\documentclass[a4paper,10pt,twoside]{cpc-hepnp}
\usepackage{CJK,upgreek,fancyhdr}
\usepackage{multicol}
\usepackage{graphicx}
\usepackage{booktabs}
\usepackage{amssymb,bm,mathrsfs,bbm,amscd}
\usepackage[tbtags]{amsmath}
\usepackage{lastpage}

\begin{document}
\begin{CJK*}{GBK}{song}

\fancyhead[c]{\small Chinese Physics C~~~Vol. xx, No. x (201x) xxxxxx}
\fancyfoot[C]{\small 010201-\thepage}

\footnotetext[0]{Received 15 October 2015}

\title{Global $\alpha$-decay study based on the mass table of the relativistic continuum Hartree-Bogoliubov theory\thanks{Supported by Major State 973 Program of China (2013CB834400), National Natural Science Foundation of China (11175002, 11335002, 11375015, 11461141002), Research Fund for the Doctoral Program of Higher Education (20110001110087) and National Undergraduate Innovation Training Programs of Peking University.}}

\author{%
      Lin-Feng ZHANG $^{1,2}$
\quad Xue-Wei XIA $^{3:1)}$\email{xuewei.xia@buaa.edu.cn}
}
\maketitle

\address{%
$^1$ State Key Laboratory of Nuclear Physics and Technology, School of Physics, Peking University, Beijing 100871, China\\
$^2$ Yuanpei College, Peking University, Beijing 100871, China\\
$^3$ School of Physics and Nuclear Energy Engineering, Beihang University, Beijing 100191, China\\
}

\begin{abstract}
The $\alpha$-decay energies ($Q_\alpha$) are systematically investigated with the nuclear masses for $10 \leq Z \leq 120$ isotopes obtained by the relativistic continuum Hartree-Bogoliubov (RCHB) theory with the covariant density functional PC-PK1, and compared with available experimental values. It is found that the $\alpha$-decay energies deduced from the RCHB results present similar pattern as those from available experiments. Owing to the large predicted $Q_\alpha$ values ($\geq$ 4 MeV), many undiscovered heavy nuclei in the proton-rich side and super-heavy nuclei may have large possibilities for $\alpha$-decay. The influence of nuclear shell structure on $\alpha$-decay energies is also analysed.
\end{abstract}

\begin{keyword}
$\alpha$-decay, RCHB theory, mass table, shell structure
\end{keyword}

\begin{pacs}
21.10.Dr, 21.60.Jz, 23.60.+e
\end{pacs}

\footnotetext[0]{\hspace*{-3mm}\raisebox{0.3ex}{$\scriptstyle\copyright$}2013
Chinese Physical Society and the Institute of High Energy Physics
of the Chinese Academy of Sciences and the Institute
of Modern Physics of the Chinese Academy of Sciences and IOP Publishing Ltd}%

\begin{multicols}{2}

\section{Introduction}

The discovery of nuclear radioactivities one century ago may be considered as the beginning of nuclear physics~\cite{Becquerel1896CRASP}. Up to now, several nuclear decay modes have been observed experimentally, such as $\alpha$-decay, $\beta$-decay, orbital electron capture, spontaneous fission, proton decay and neutron decay.~\cite{Sobiczewski2007}. One of the most important decay modes is $\alpha$-decay, the investigation on which plays an essential role in exploring the nuclear structure, especially for nuclei in the heavy and super-heavy nuclear regions~\cite{Sobiczewski2007}, and it's also a hot topic of current research.

One crucial characteristic quantity of an $\alpha$-emitter is the $\alpha$-decay energy $Q_\alpha$, which is defined as
\begin{eqnarray}\label{eq1}
  Q_{\alpha}=E_B(Z-2,N-2)+E_B(2,2)-E_B(Z,N),
\end{eqnarray}
where $E_B(Z,N)$ is the binding energy for the nucleus with proton number $Z$ and neutron number $N$. With the $\alpha$-decay energies, a number of empirical formula have been proposed for the half-lives of nuclei~\cite{ViolaJr1966741, Hatsukawa1990PhysRevC.42.674, Brown1992PhysRevC.46.811, royer2000alpha, dong2005new}. One of the necessary conditions for a nucleus to spontaneously emit an $\alpha$-particle is that the $\alpha$-decay energy $Q_{\alpha}$ must be positive. Consequently, in order to investigate the $\alpha$-decay energies, the precise nuclear masses are needed. Experimentally, nuclear masses of more than $2000$ nuclei have been measured thanks to the application of cyclotron, storage ring and penning trap facilities~\cite{Wang2012CPC}. However, $\alpha$-decay is also expected to happen in the large unknown region of nuclear chart, which is still beyond the experimental capability in the foreseeable future. Therefore, a systematical investigation for $\alpha$-decay energies has to rely on robust theoretical nuclear mass models.

Theoretical investigations on nuclear masses can be classified into the following two categories. The first consists of macroscopic-microscopic models, such as Liquid drop model (LDM)~\cite{Weizsacker1935}, finite-range droplet model (FRDM)~\cite{Moller1995}, extended Thomas-Fermi plus Strutinsky integral with shell quenching (ETFSI-Q) model~\cite{Pearson1996} and Weizsacker-Skyrme mass model (WS)~\cite{PhysRevC.82.044304,wang2014surface,PhysRevC.84.014333}. The second is composed of microscopic models, for example, the Hartree-Fock-Bogoliubov (HFB) model~\cite{Goriely2009PRL, Goriely2009PRL102.242501, Goriely2010} based on density functional theory (DFT), which is believed to have a reliable extrapolation to the unknown regions.

Nowadays, the covariant density functional theory (CDFT) has attracted extensive attention because of the successful description of many nuclear phenomena~\cite{Serot1986relativistic, Ring1996PPNP, Vretenar2005reports, Meng2006PPNP, Meng2011covariant, Nikvsic2011PPNP, Meng2013frontier}. It can provide a natural inclusion of the nucleon spin degree of freedom, resulting in the nuclear spin-orbit potential that emerges automatically with the empirical strength in a covariant way. It provides a new saturation mechanism for nuclear matter~\cite{Brockman1990PhysRevC.42.1965}, reproduces well the measured isotopic shifts in the Pb region~\cite{Sharma1993PLB}, reveals more naturally the origin of the pseudospin symmetry~\cite{Arima1969PLB, Hecht1969NPA} as a relativistic symmetry ~\cite{Ginocchio1997PhysRevLett.78.436, Meng1998PhysRevC.58.R628, Meng1999PhysRevC.59.154, Ginocchio2004PhysRevC.69.034303, Long2006PLB, Liang2011PhysRevC.83.041301, Lu2012PhysRevLett.109.072501, Guo2012PhysRevC.85.021302, Lu2013PhysRevC.88.024323,Guo2014PhysRevLett.112.062502, Liang2015report}, and predicts the spin symmetry in the anti-nucleon spectrum \cite{Zhou2003PhysRevLett.91.262501, Liang2010EJPA}. It can also include the nuclear magnetism~\cite{Koepf1989NPA}, i.e., a consistent description of currents and time-odd fields, which plays a crucial role in the nuclear magnetic moments ~\cite{Yao2006PhysRevC.74.024307, Arima2011scichina, liJ2011PTP, wei2012relativistic}, nuclear rotations~\cite{Afanasjev2000NPA, Zhao2012PhysRevC.85.054310, Zhao2011PhysRevLett.107.122501, Zhao2011PLB}. The CDFT is a reliable and useful model for nuclear structure study in the whole nuclear chart.

The first CDFT mass table calculated 2000 even-even nuclei with $8 \leq Z \leq 120$~\cite{Hirata1997NPA}, but without treating pairing correlations. Later, using Bardeen-Cooper-Schrieffer (BCS) method, the ground-state properties of 1315 even-even nuclei with $10 \leq Z \leq 98$ were calculated~\cite{Lalazissis1999ADNDT}. In 2005, by employing the state-dependent BCS method with a delta pairing force, the first systematic study of the ground-state properties for about 7000 nuclei was performed~\cite{Geng01042005}. More recently, the RHB framework is used for a systematic study of ground state properties of all even-even nuclei from the proton to neutron drip line ~\cite{Afanasjev2013PLB, Agbemava2014}.

It is widely considered that pairing correlation has a critical influence on open shell nuclei~\cite{Meng2006PPNP}. Among the methods in dealing with pairing correlation, Bogoliubov quasiparticle transformation is generally used for exotic nuclei which can include the continuum appropriately when treated in coordinate representation~\cite{Meng2002Phys.Rev.C041302}. As an extension of the relativistic mean field and the Bogoliubov transformation in the coordinate representation, relativistic continuum Hartree-Bogoliubov (RCHB) theory provides a fully self-consistent description of both the continuum and the bound states as well as the coupling between them~\cite{Meng1996Phys.Rev.Lett.3963, Meng1998Phys.Rev.Lett.460, Meng1998Nucl.Phys.A3}. The halo in $^{11}\rm{Li}$ has been described~\cite{Meng1996Phys.Rev.Lett.3963, Meng1998Nucl.Phys.A3} and the giant halos in light and medium-heavy nuclei were predicted~\cite{Meng1998Phys.Rev.Lett.460, Meng2002Phys.Rev.C041302, zhang2003SIC}. In addition, the generalization to the odd-nucleon system~\cite{Meng1998Phys.Lett.B1,Meng2002209} and deformed nuclei~\cite{Zhou2010PRC, Li2012PhysRevC.85.024312, Chen2012PhysRevC.85.067301} were developed.

To investigate the impact of the continuum for the nuclear chart, the RCHB theory is used to systematically calculate nuclear masses for $8 \leq Z \leq 120$ isotopes by assuming spherical symmetry. Taking the nuclear chart ranging from O to Ti as an example, the influence of continuum on nucleon drip-lines has been investigated in Ref.~\cite{Qu2013CPC}. It shows that although the proton drip-lines predicted with various mass models, such as FRDM~\cite{Moller1995}, WS3~\cite{PhysRevC.84.014333}, HFB-21~\cite{Goriely2010} and TMA~\cite{Geng01042005} are roughly the same and basically agree with the observation, the neutron drip-line predicted by RCHB theory with the covariant density functional PC-PK1 is extended further neutron-rich than other mass models due to the continuum couplings. Therefore, it is interesting to systematically study the nuclear ground-state properties , such as nuclear mass and radius, by using the mass table provided by RCHB theory. Meanwhile, it is also allowed to systematically study nuclear decay modes related to the nuclear masses.\par

In this paper, the $\alpha$-decay energies will be systematically investigated with the nuclear masses for $8 \leq Z \leq 120$ isotopes provided by RCHB theory with the covariant density functional PC-PK1~\cite{Xia2015}. The RCHB results are compared with available experimental values. The influence of nuclear shell structure on $\alpha$-decay energies is also investigated.

\section{Theoretical Framework}

Starting from the effective Lagrangian density
\begin{eqnarray}\label{EQ:LAG}
  {\cal L} ={\cal L}^{\rm{free}}+{\cal L}^{\rm{4f}}+{\cal L}^{\rm{hot}}+{\cal L}^{\rm{der}}+{\cal L}^{\rm{em}},
\end{eqnarray}
where
 \begin{eqnarray}
 {\cal L}^{\rm{free}}&=&\bar\psi(i\gamma_\mu\partial^\mu-m)\psi\\
{\cal L}^{\rm{4f}~~} &=&-\frac{1}{2}\alpha_S(\bar\psi\psi)(\bar\psi\psi)
-\frac{1}{2}\alpha_V(\bar\psi\gamma_\mu\psi)(\bar\psi\gamma^\mu\psi)\nonumber\\
& &
-\frac{1}{2}\alpha_{TV}(\bar\psi\vec{\tau}\gamma_\mu\psi)(\bar\psi\vec{\tau}\gamma^\mu\psi)\\
{\cal L}^{\rm{hot}}&=& -\frac{1}{3}\beta_S(\bar\psi\psi)^3-\frac{1}{4}\gamma_V[(\bar\psi\gamma_\mu\psi)(\bar\psi\gamma^\mu\psi)]^2\nonumber\\
& &-\frac{1}{4}\gamma_S(\bar\psi\psi)^4\\
   {\cal L}^{\rm{der}}&=&-\frac{1}{2}\delta_S\partial_\nu(\bar\psi\psi)\partial^\nu(\bar\psi\psi)
-\frac{1}{2}\delta_V\partial_\nu(\bar\psi\gamma_\mu\psi)\partial^\nu(\bar\psi\gamma^\mu\psi)\nonumber\\
& &
-\frac{1}{2}\delta_{TV}\partial_\nu(\bar\psi\vec\tau\gamma_\mu\psi)\partial^\nu(\bar\psi\vec\tau\gamma_\mu\psi)\\
  {\cal L}^{\rm{em}}&=&-\frac{1}{4}F^{\mu\nu}F_{\mu\nu}-e\frac{1-\tau_3}{2}\bar\psi\gamma^\mu\psi A_\mu,
 \end{eqnarray}
one can derive the RHB equation for the nucleons~\cite{Kucharek&Ring1991ZPA},
\begin{eqnarray}
\left(\begin{array}{cc}
h_D-\lambda & \Delta\\
-\Delta^*& -h^*_D+\lambda
\end{array}\right)
\left(\begin{array}{c}
U_k\\
V_k
\end{array}\right)
=E_k\left(\begin{array}{c}
U_k\\
V_k
\end{array}\right),
\end{eqnarray}
where
\begin{eqnarray}
 h_D(\mathbf{r}) =\bm{\alpha} \cdot \mathbf{p} + V(\mathbf{r}) + \beta (M + S(\mathbf{r})),
\end{eqnarray}
and find the solution self-consistently. With the spherical symmetry, the RCHB theory solves the RHB equations in coordinate space. For the detailed formalism and numerical techniques, see Ref.~\cite{Meng1998Nucl.Phys.A3} and references therein. In the present calculations, we follow the procedures in Refs.~\cite{Meng1998Nucl.Phys.A3, Meng1998Phys.Lett.B1} and solve the RCHB equations in a box with the size $R$=20 fm and with a step of 0.1 fm. In addition, we use the density functional PC-PK1 \cite{Zhao2010Phys.Rev.C054319} for particle-hole channel, for particle-particle channel, the density-dependent delta pairing force
\begin{eqnarray}
V(\mathbf{r}_1,\mathbf{r}_2)=V_0\delta(\mathbf{r}_1-\mathbf{r}_2)\frac{1}{4}[1-\bm{\sigma}_1\bm{\sigma}_2](1-\frac{\rho(\bm{r}_1)}{\rho_0}),
\end{eqnarray}
  is employed. In equation (10) the saturation density $\rho_{0}=0.152~\rm{fm^{-3}}$ and the pairing strength $V_0=685.0~\rm{MeV\cdot fm^{3}}$ is fixed by reproducing experimental odd-even mass differences of  $Z$ = 20, 50, 78, 92 isotope chains and $N$ = 20, 50, 78, 92 isotone chains, respectively. The contribution from the continuum is restricted within a cutoff energy $E_{\rm{cut}}=100~\rm{MeV}$ and cutoff angular momentum $j_{\rm{max}}=\frac{19}{2}\hbar$.

\section{Results and Discussion}

By using the binding energies provided in the RCHB theory with the density functional PC-PK1~\cite{Zhao2010Phys.Rev.C054319}, the $Q_\alpha$ values of 9035 predicted bound nuclei with $10 \leq Z \leq 120$~\cite{Xia2015} are obtained with Eq.~(1). It is found that the values of 3703 nuclei, plotted with different colors in Fig.~\ref{fig1}(a), are positive. Among these nuclei, the $Q_\alpha$ values of 1629 nuclei are less than 4 MeV,  of 1299 nuclei are within 4-8 MeV, of 734 nuclei are within 8-12 MeV, and of 41 nuclei are larger than 12 MeV.  Several systematical features can be found from Fig.~\ref{fig1}(a):
1) From a global view, most nuclei with positive $Q_\alpha$ value are located in the upper-left side of the nuclear chart;
2) For a given isotope chain, $Q_\alpha$ generally decreases with the increase of neutron number $N$;
3) For a given isotone chain, $Q_\alpha$ generally increases with the increase of proton number $Z$;
4) The $Q_\alpha$ value can be greatly influenced by the existed shell structure, which can be clearly seen from the sudden increase of $Q_\alpha$ when $Z$ or $N$ cross the magic numbers 28, 50, 82 and 126;
5) The lightest nucleus predicted to have positive $Q_\alpha$ value in RCHB mass table is $^{20}$Ne ($Q_\alpha$=0.14 MeV),
and then several $Z\approx N$ nuclei in the $A\sim60$ mass region;
6) Nuclei with very large $Q_\alpha$ values ($>$4 MeV) are mostly the heavy or superheavy neutron-deficient nuclei;
7) Remarkably, in the superheavy mass region around $Z\sim$120, positive $Q_\alpha$ can be even extended to neutron-rich region.

\end{multicols}
\ruleup
\begin{center}
\includegraphics[width=12cm]{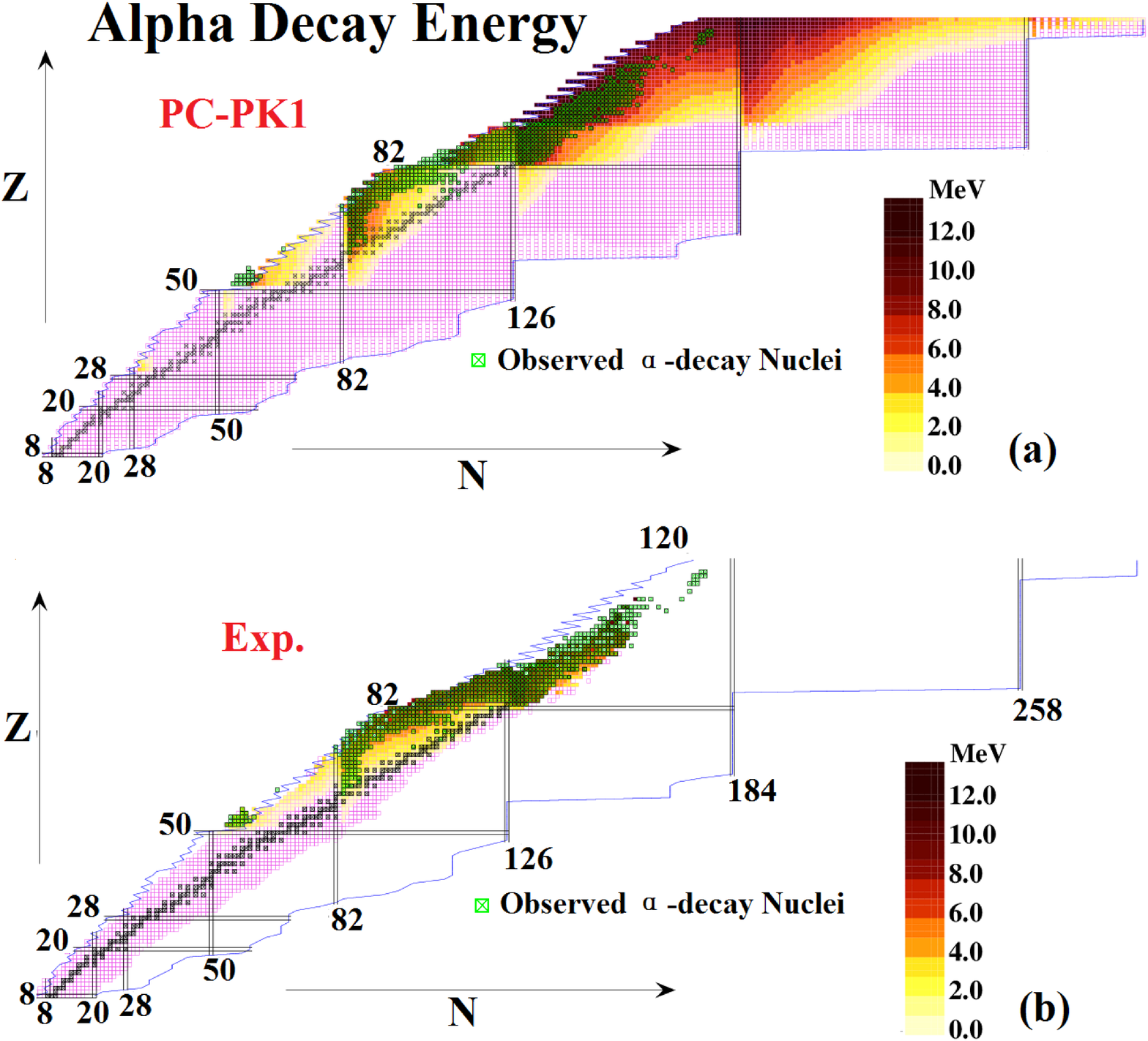}
\figcaption{\label{fig1} (Color online) $\alpha$-decay energies $Q_{\alpha}$ for nuclei with $10 \leq Z \leq 120$, provided by (a) RCHB theory with the density functional PC-PK1 \cite{Zhao2010Phys.Rev.C054319} and (b) available experimental values \cite{Audi2012CPC}. Blue lines are proton and neutron drip-lines predicted by the RCHB theory. The nuclei predicted to be bound in present work and observed experimentally are represented as the squares in panel (a) and (b), respectively. Furthermore, 719 nuclei observed experimentally with the radioactivity of $\alpha$-decay are marked with green crosses.}
\end{center}
\ruledown
\begin{multicols}{2}

Analogously, the experimental $Q_\alpha$ values are obtained with Eq.~(1) from the evaluated mass data in AME2012 mass table~\cite{Audi2012CPC} and those of 1067 nuclei, plotted in Fig.~1(b), are found to be positive. In particular, there are 719 nuclei observed experimentally with the radioactivity of $\alpha$-decay at present~\cite{http://www.nndc.bnl.gov/}, being marked with green crosses in Figs.~\ref{fig1}(a) and~\ref{fig1}(b). It should be emphasized that it is not the aim to compare theoretical and experimental $Q_\alpha$ values detailedly, but rather to investigate the schematic evolution of the $\alpha$-decay energies.

Although the general features of $\alpha$-decay have been well known, for completeness, the following remarks are noted here from Fig.~\ref{fig1}(b):
1)The lightest nucleus presently found to have the radioactivity of $\alpha$-decay is $^{105}$Te~\cite{http://www.nndc.bnl.gov/}, and $A\sim$100 marks the lightest mass region with the radioactivity of $\alpha$-decay;
2) Globally, $\alpha$-decay is mainly observed in the neutron deficient nuclei with $N\geq84$;
3) In the region $N\ge 126$, most of the nucleus are found to have the radioactivity of $\alpha$-decay, and particularly, almost all those observed superheavy nuclei ($Z\geq110$) have $\alpha$-decay radioactivity;
4) Due to the Coulomb barrier and the competition of other decay modes, not all nuclei with positive $Q_\alpha$ values are observed to have $\alpha$-decay radioactivity in ground state. Among the 719 nuclei observed with $\alpha$-decay radioactivity, $^{187}\rm{Re}$ is the one with smallest $Q_\alpha$ value (1.66 MeV), and about 70 percent are with $Q_\alpha$ values larger than 4.0 MeV;
5) There are 68 nuclei with $Q_\alpha$ values larger than 4.0 MeV being not observed the $\alpha$-radioactivity in ground state,
which, however, have been found to decay by other modes, such as $\beta^+$, EC, $\beta^-$ decays or spontaneous fission.

When comparing the two panels of Fig.~\ref{fig1}, the following features can be addressed:
1) It can be found that although spherical symmetry is assumed, the positive $\alpha$-decay energies deduced from the RCHB results still present similar pattern as those from available experimental values;
2) In Fig.~1(a) most of the neutron-deficient nuclei in the heavy and superheavy regions are predicted to have $Q_{\alpha}$ values larger than 4 MeV, therefore it is expected these nuclei should have large possibilities for $\alpha$-decay, in consistent with the region of observed nuclei with $\alpha$-radioactivity;
3) In the unexplored superheavy nuclear region with $N > 184$ and $Z > 92$, there exhibits a triangle-like region for nuclei with $Q_{\alpha}>$ 4 MeV or even $>$ 10 MeV, indicating the possibility of $\alpha$-radioactivity for these neutron rich nuclei;
4) It is also noted that 74 exotic nuclei proved to be bound in experiments, mainly in the neutron-deficient region near $Z$=50 and $Z$=82, locate beyond the proton drip-line of RCHB mass table and are absent in the present prediction, which needs to be further examined in the future, for instance, by taking into account the deformation effect and Wigner term \cite{Wigner_PhysRev.51.106}.

\end{multicols}
\ruleup
\begin{center}
\includegraphics[width=12cm]{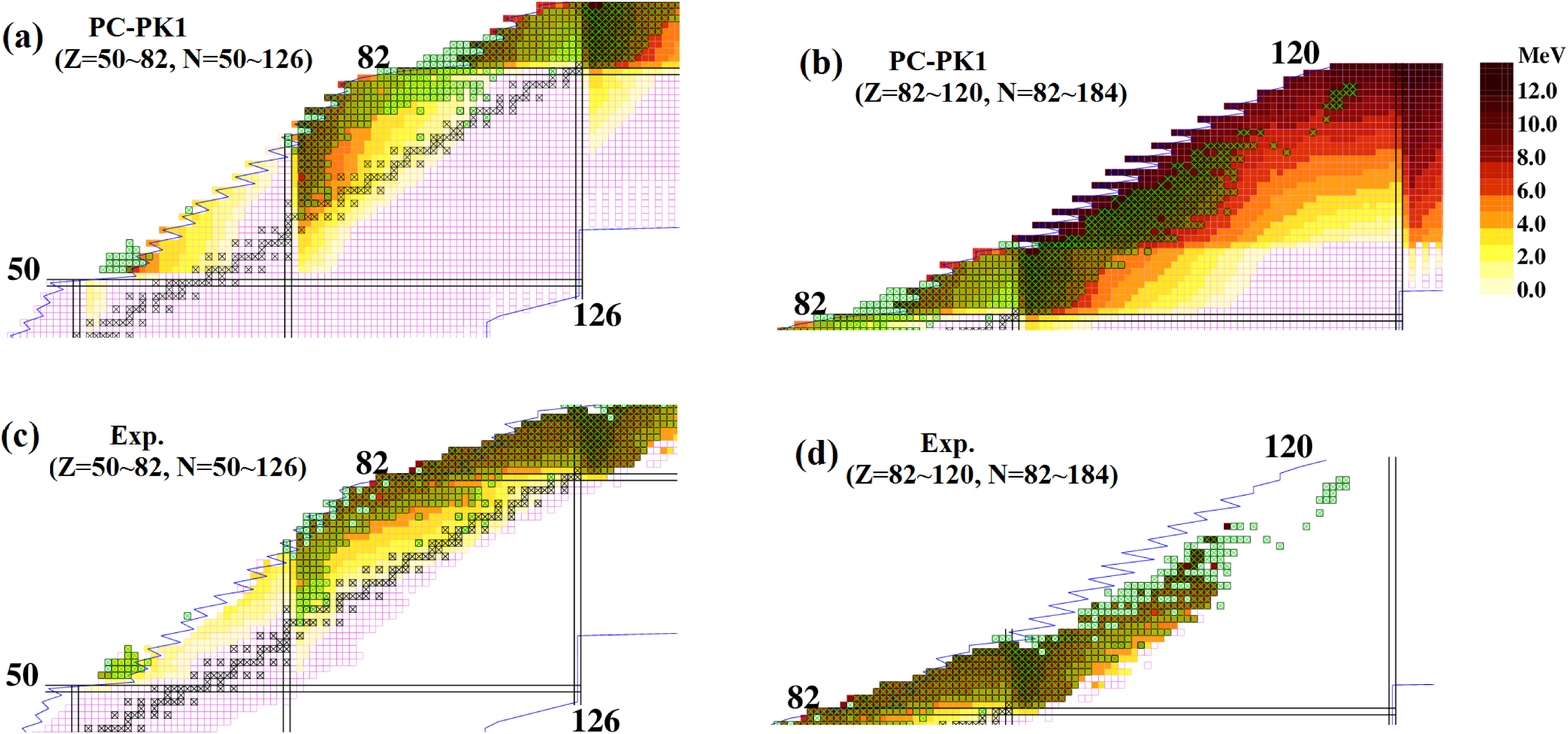}
\figcaption{\label{fig2} (Color online) Comparison between the calculated and experimental $Q_\alpha$ values in the nuclear regions of ($Z$ = 50 $\sim$ 82, $N$ = 50 $\sim$ 126), and ($Z$ = 82 $\sim$ 120, $N$ = 82 $\sim$ 184).}\label{fig2}
\end{center}
\ruledown
\begin{multicols}{2}

To closely inspect the evolution of $\alpha$-decay energy with proton and neutron numbers, comparison between the calculated and experimental $Q_\alpha$ values in two specific mass regions ($Z$ = 50 $\sim$ 82, $N$ = 50 $\sim$ 126) and ($Z$ = 82 $\sim$ 120, $N$ = 82 $\sim$ 184) is given in Fig. 2. The shell effect can be clearly seen here. Taking the mass region around the doubly magic numbers $Z$=82 and $N$=126 as an example, the $Q_\alpha$ value of a nucleus with $Z>82$ and $N>126$, $Q_\alpha(Z>82, N>126)$, is systematically larger than $Q_\alpha(Z>82, N<126)$, and then $Q_\alpha(Z<82, N>126)$, and then $Q_\alpha(Z<82, N<126)$, which owes to the partitioning of the magic numbers $Z$=82 and $N$=126. As a magic nucleus provides more stability, according to Eq.~(1), the $Q_\alpha$ value of a magic nucleus is much smaller than that of the nucleus with two more protons or neutrons.

\begin{center}
\includegraphics[width=6cm]{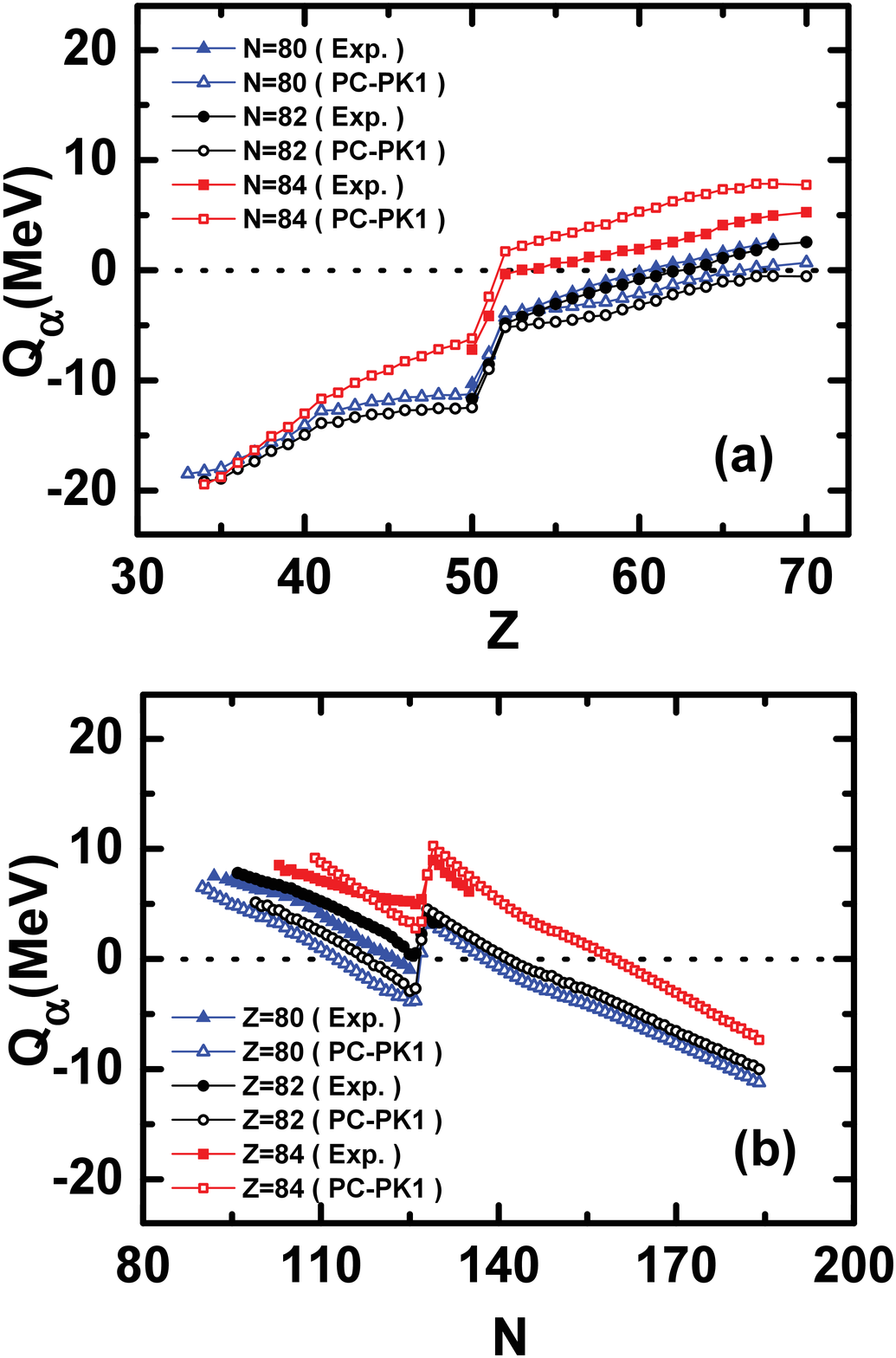}
\figcaption{\label{fig3} (Color online) $\alpha$-decay energies $Q_{\alpha}$ for (a) $N$=80, 82, 84 isotone chains; (b) $Z$=80, 82, 84 isotope chains.}
\end{center}

\begin{center}
\includegraphics[width=6cm]{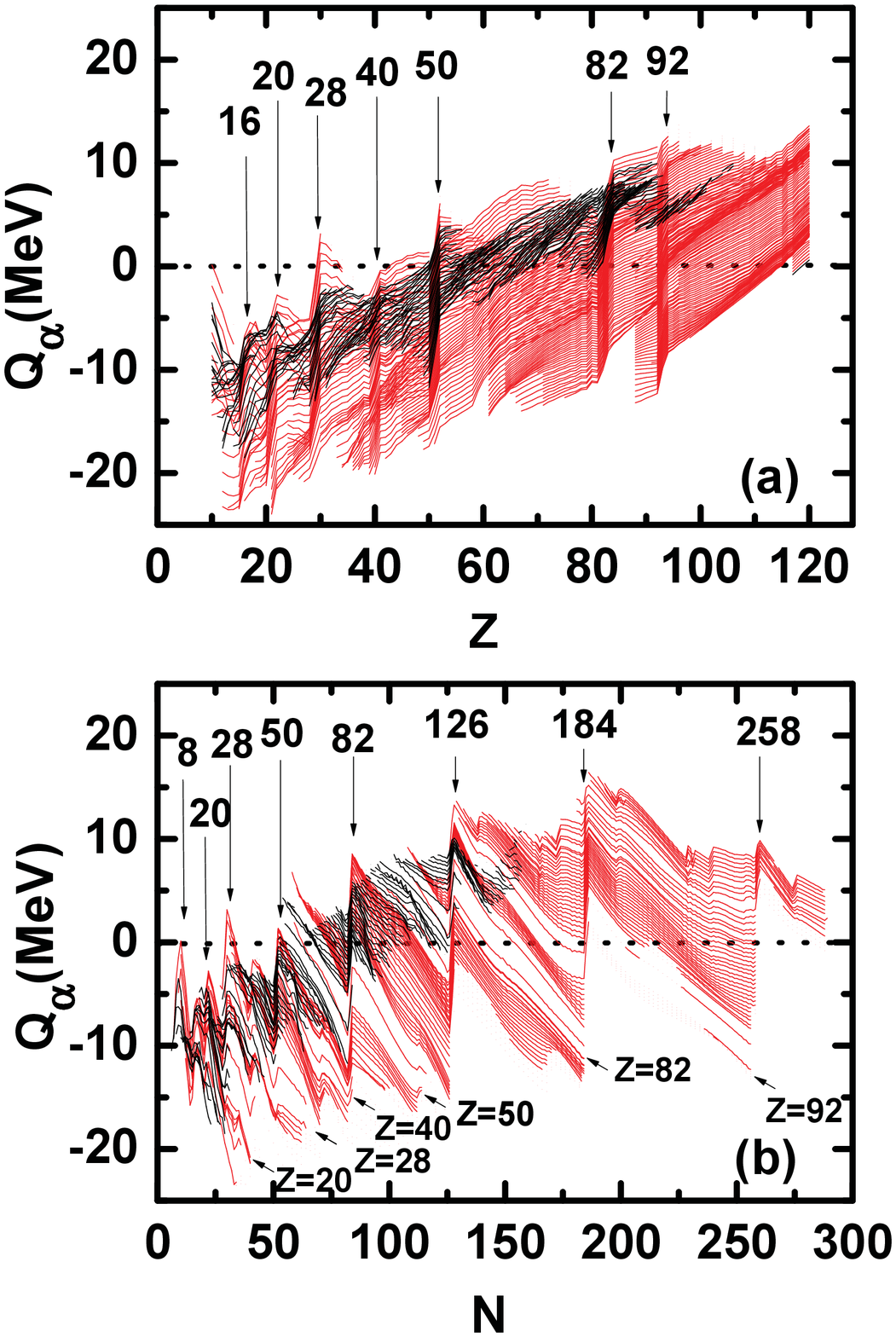}
\figcaption{\label{fig4} (Color online) $\alpha$-decay energies $Q_{\alpha}$ as functions of proton(a) and neutron(b) numbers. The red lines represent the results of the RCHB theory with the covariant density functional PC-PK1; the black lines denote the experimental values from Ref. \cite{Audi2012CPC}. }
\end{center}

The calculated and experimental $\alpha$-decay energies $Q_{\alpha}$ for $N$ ($Z$) = 80, 82, 84 isotone (isotope) chains are respectively shown in Fig. 3(a) and (b), and good agreement between them can be easily found.
The influences of shell closure on the decay energies $Q_\alpha$ are illustrated in two aspects. First, as shown in the panels, a sudden increase of $Q_\alpha$ exists at the magic number $Z = 52$ ($N=128$). Second, by comparing the three isotone (isotope) chains $N$ ($Z$) = 80, 82, 84, both the calculated and experimental values of $Q_{\alpha}$ at $N$ ($Z$) = 84 are clearly larger than the corresponding values at $N$ ($Z$) = 80, 82.

Therefore, the sudden increase of the $Q_\alpha$ value along $Z$ or $N$ can be used as a probe for possible shell closures. In Fig. 4, the theoretical and experimental $\alpha$-decay energies $Q_\alpha$ as functions of $Z$ ($N$) are plotted for all the isotopic (isotonic) chains. As shown in Fig. 4, the sudden increases exist at the traditional proton magic numbers $Z$ = 20, 28, 50, 82, and the neutron magic numbers $N$ = 8, 20, 28, 50, 82, 126. Similar sudden increases of the theoretical $Q_\alpha$ value can also be clearly found at $Z$= 16, 40, 92, $N$= 184, 258, where $Z$ = 16 has been proved as an magic number close to the neutron drip line \cite{Ozawa2000Phys.Rev.Lett.5493,Kanungo}; $Z$= 40 is generally considered as a sub-shell; $Z$ = 92 is considered as a pseudo-shell in the relativistic mean field calculations; and $N$ = 184, 258 are possibly the new magic numbers in the superheavy mass region, as suggested in the previous RCHB calculations with a number of effective interactions NL1, NL3, NLSH, TM1, TW99, DD-ME1, PK1 and PK1R~\cite{Zhang2005NPA}.

\section{Summary and Perspective}

In conclusion, the $\alpha$-decay energies with RCHB mass table are systematically studied. The $Q_{\alpha}$ values calculated by RCHB theory with the covariant density functional PC-PK1 agree well with experimental values. It is shown by available experimental values that $\alpha$-decay is mainly observed in the proton-rich and heavy nuclear regions, and the values of observed $Q_{\alpha}^{exp.}$ for most $\alpha$-decay nuclei are larger than 4 MeV. In addition, illustrated by calculated results, most of the decay energies $Q_{\alpha}$ predicted in the proton-rich heavy and super-heavy nuclear regions are larger than 4 MeV, which may indicate the large possibility for them to have $\alpha$-decay. By plotting $\alpha$-decay energies $Q_{\alpha}$ for $N(Z) =$ 80, 82, 84 isotone chains (isotope chains) calculated by RCHB theory and experimental values, the influences of shell effect on $\alpha$-decay energies are also investigated in detail. It is found that an abrupt change of $Q_{\alpha}$ exists when crossing over each magic number. Furthermore, by plotting $\alpha$-decay energies with proton number $Z$ and neutron number $N$ respectively,
the traditional magic numbers are reproduced by the sudden increase of $Q_{\alpha}$ there, and the possible new magic numbers $N =$ 184 and 258 are predicted.

In the future, RCHB mass table can be used to calculate decay energies of C, O clusters, and study them in a much similar way. In addition, deformed relativistic Hartree-Bogoliubov theory in continuum (DRHBc) can be used to study $\alpha$-decay energies, and investigate the influence of deformation on the $\alpha$-decay energies.

\acknowledgments{Precious supervision from Jie Meng and Shuangquan Zhang and fruitful discussions with Ying Chen, Yeunhwan Lim, He Liu, Xiaoying Qu, Ik Jae Shin, Pengwei Zhao and Youngman Kim are acknowledged.}

\end{multicols}

\vspace{-1mm}
\centerline{\rule{80mm}{0.1pt}}
\vspace{2mm}

\begin{multicols}{2}

\end{multicols}

\clearpage
\end{CJK*}
\end{document}